\documentclass[conference, a4paper]{IEEEtran}
\IEEEoverridecommandlockouts

\ifCLASSINFOpdf
  \usepackage[pdftex]{graphicx}
  \DeclareGraphicsExtensions{.pdf,.jpeg,.png}
\else
\fi

\usepackage{cite}
\usepackage{amsmath,amssymb,amsfonts}
\usepackage{algorithmic}
\usepackage{graphicx}
\usepackage{textcomp}
\usepackage[left=1.4cm, right=1.4cm, top=1.7cm]{geometry}
\usepackage{caption}
\usepackage{anyfontsize}
\usepackage{url}
\usepackage{xcolor} 

\makeatletter
\newcommand*\titleheader[1]{\gdef\@titleheader{#1}}
\AtBeginDocument{%
\let\st@red@title\@title
\def\@title{%
\bgroup\normalfont\normalsize\centering\@titleheader\par\egroup
\vskip0.2em\st@red@title}
}
\makeatother

\makeatletter
\renewcommand{\fnum@figure}{Figure \thefigure}
\makeatother

\title{ {Structural predictability of large-scale aircraft interaction networks} \\
\thanks{This project has received funding from the European Research Council (ERC) under the European Union's Horizon 2020 research and innovation programme (grant agreement No 851255). This work was partially supported by the Mar\'ia de Maeztu project CEX2021-001164-M funded by the MICIU/AEI/10.13039/501100011033. R. L.-M. acknowledges support from the Spanish Ministry of Science, Innovation and Universities through the grant FPU22/03765.}
\vspace{0.5cm}
}

\titleheader{First US-Europe Air Transportation Research and Development Symposium (ATRDS2025)}

\author{\IEEEauthorblockN{Raúl López-Martín and Massimiliano Zanin}
\IEEEauthorblockA{Instituto de Física Interdisciplinar y Sistemas Complejos CSIC-UIB \\
Campus Universitat de les Illes Balears \\
Palma de Mallorca, Spain \\
\{raullopez, mzanin\}@ifisc.uib-csic.es }

}

\renewcommand\IEEEkeywordsname{Keywords}
\IEEEaftertitletext{\vspace{-1\baselineskip}}

\begin{document}

\maketitle

\noindent \begin{abstract}
Complex network theory has recently been proposed as a promising tool for characterising interactions between aircraft, and their downstream effects. We here explore the problem of networks' topological predictability, i.e. the dependence of their structure on the traffic level, but the apparent absence of significant inter-day variability. By considering smaller spatial scales, we show that the sub-networks corresponding to individual FIRs are highly heterogeneous and of low predictability; this is nevertheless modulated by the structure of airways, and specifically by the complexity in airspace usage. We further discuss initial results of the evolution of such properties across multiple spatial scales; and draw conclusions on the operational implications, specifically on efforts to limit downstream effects.
\end{abstract}

\vspace{0.3cm}


\section{Introduction}

While separation assurance has classically been focused on maintaining a safe distance between two (or more) aircraft, in recent years an increasing attention has been devoted to expand this horizon to include downstream effects. In other words, one is not only interested in flights that will take part in conflicts, but also how the resolution of these can cascade into additional conflicts. Beyond the clear impact in the safety of operations, this is also relevant towards new operational concepts of higher efficiency, as e.g. flight-centric \cite{gerdes2018dynamic}, flow-centric \cite{schultz2023introduction} and free-flight operations \cite{hoekstra2002designing}. The task is nevertheless a challenging one, as such derived conflicts may depend both on the choices made by air traffic controllers and on local conditions, e.g. winds; in other words, and as ubiquitous in air traffic, on trajectories' uncertainty.

A recently proposed solution involves the use of complex networks to describe the structure created by aircraft interactions within an airspace \cite{lopez2023, lopez2024}. Such analysis involves, firstly, the identification of interactions, i.e. instances of reduced separation (not necessarily a Loss of Separation) that require an increase in attention or resolution maneuvers. Secondly, a network is reconstructed, where each flight is represented via a node, and two of them are connected via an edge when they take part in an interaction. These networks can then be described through topological metrics, i.e. measurements quantifying the presence of specific connectivity structures. This approach allows to get a picture of how interactions could potentially propagate beyond pairs of flights, while being conceptually simple and computationally parsimonious.

In a previous work \cite{lopez2024} this approach has been applied to real European traffic, with the surprising result that the interaction structure strongly depends on the number of flights and on the day of the week, but shows limited variability otherwise. In other words, suppose one considers two Mondays of two different years; once compensated for the traffic level, the underlying interaction network will have the same properties, irrespective of the specific events that may have happened. Conversely, this implies that actions taken to modify such network structure will have a limited impact. In other words, these interaction networks are both predictable and rigid.

The aim of this work is to shed light on this problem, and specifically on what are the factors affecting the predictability (or rigidity) of the interaction networks. This is tackled through a spatial multiscale analysis of these European interaction networks across different FIRs. After introducing the operational data here considered (Sec. \ref{sec:operational_data}), we revisit how interaction networks are reconstructed and how their topology evolves through time (Sec. \ref{sec:global}). We next propose a FIR-based analysis of the same networks (Sec. \ref{sec:FIR_based}), establishing connections with the complexity of the underlying airspace usage; and a multiscale extension of the same (Sec. \ref{sec:multiscale}). We conclude by discussing the relevance of these results, and proposing some hypotheses to be further tested.

\section{Operational data}
\label{sec:operational_data}

Data used in this work come from the EUROCONTROL's R\&D Data Archive, a public repository of historical flights made available for research purposes and freely accessible at \url{https://www.eurocontrol.int/dashboard/rnd-data-archive}. Data are limited at source to four months (i.e. March, June, September and December) of each year; we have here considered the time span from March $1^{st}$ 2015 to March $13^{th}$ 2020, in order to avoid biases due to the COVID-19 pandemic. All available executed trajectories were cropped following an approximation of the European airspace, spanning between $-15^{\circ}$ and $30^{\circ}$ in longitude, and between $35^{\circ}$ and $70^{\circ}$ in latitude. We further used the airspace structure in there reported, specifically the structure of FIRs in each AIRAC cycle.

\section{Reconstruction and properties of the global interaction networks}
\label{sec:global}

The reconstruction of the interaction networks is based on the methodology previously presented in Refs. \cite{lopez2023, lopez2024}. For the sake of completeness, we here include a synthesis of the methodology; additional details and sensitivity analyses can be found in the aforementioned works.

Each available day is represented by an independent network; individual flights are mapped to nodes, pairwise connected when the corresponding distance falls below a separation threshold of $10$NM horizontally and $2,000$fts vertically. Links are non-directed and inherit the temporal timestamp of the moment of the interaction. Only one edge can exist between a pair of nodes, corresponding to the first occurrence of an interaction. Finally, downstream effects are represented through paths, that is, collection of nodes that are connected through chronologically-ordered links - to prevent the appearance of propagations going backwards in time. 
These networks of interactions are analysed through six classical topological metrics, i.e., measurements quantifying specific aspects of their structure. For the sake of 
completeness, a synthetic description of their meaning is reported in Tab. \ref{tab:metrics_def}; additional information can also be found in Refs. \cite{lopez2023, lopez2024}.

\begin{table}[tb]
\caption{Topological metrics considered in this study, with their description and relevant references. }
\begin{center}
\begin{tabular}{|p{1.45cm}|p{5.7cm}|p{0.45cm}|}
\hline
\textbf{Metric}& \textbf{Definition} & \textbf{Refs.} \\\hline
Degree entropy & Heterogeneity of the number of connections of nodes. & \cite{wang2006entropy} \\\hline
Isolated nodes & Normalised number of nodes with no interactions. & \cite{costa2007characterization} \\\hline
Weak giant cluster size & Size of the largest group of nodes which can pairwise (directly or otherwise) interact. & \cite{costa2007characterization} \\\hline
Efficiency & Quantification of how easily downstream can occur, through the average inverse of distance between pairs of nodes. & \cite{latora2001efficient} \\\hline
Mean 4 reachability & Normalised number of nodes that can be reached from a starting one through paths of length four or less. & \cite{costa2007characterization} \\\hline
Reachability modularity & Metric assessing nodes' organisation in communities, i.e. groups of nodes strongly connected between them and loosely connected with others. & \cite{fortunato2010community} \cite{fortunato2016community} \\\hline
\end{tabular}
\label{tab:metrics_def}
\end{center}
\end{table}

\begin{figure}[tb]
\centering
\includegraphics[width=\linewidth]{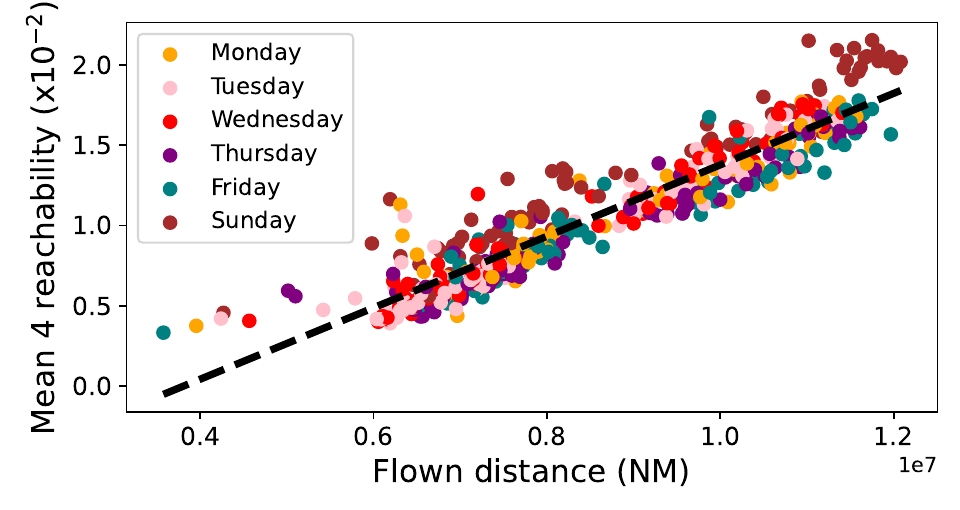}
\caption{Mean 4 reachability of the interaction networks as a function of the daily total flown distance. Each point represents one day, and its colour the day of the week (see legend). The dashed black line represents the best linear fit.}
\label{fig:metric_vs_distance_total_network}
\end{figure}

As previously introduced, the properties of the interaction networks reconstructed from flights over Europe were found to be largely defined by the daily traffic volume, for all days except for Saturdays.
This is confirmed by the high $R^2$ values of the linear fit between each of the six topological metrics reported in Tab.~\ref{tab:metrics_def} and the flown distance; all $R^2$ values but one are above $0.86$, see Tab. \ref{tab:metrics_vs_distance_full_netw_r2}. A visual example is reported in Fig. \ref{fig:metric_vs_distance_total_network}, depicting the evolution of the mean 4 reachability as a function of the flown distance, with the best linear fit represented by the black dashed line. The interested reader can also refer to Ref. \cite{lopez2024} for an in-depth analysis.
In other words, these results indicate that the global interaction structure is only marginally impacted by the natural variability of the system across days, as e.g. changes in aircraft's trajectories and delays due to weather events, and is hence predictable. In what follows, the $R^2$ of the linear fit between a metric and the flown distance will be called ``predictability''.

\begin{table}[b]
\caption{$R^2$ (i.e. predictability) of the best linear fit for each of the metrics vs. daily total flown distance. }
\begin{center}
\begin{tabular}{cc}
 \begin{tabular}{|l|c|}
\hline
\textbf{Metric}& $\mathbf{R^2}$ \\\hline
Degree entropy & 0.939 \\\hline
Isolated nodes & 0.864 \\\hline
Weak giant cluster s. & 0.866 \\\hline
\end{tabular}
    & 
\begin{tabular}{|l|c|}
\hline
\textbf{Metric}& $\mathbf{R^2}$ \\\hline
Efficiency & 0.911 \\\hline
Mean 4 reachability & 0.882 \\\hline
Reachability mod. & 0.749 \\\hline
\end{tabular}

\end{tabular}
\label{tab:metrics_vs_distance_full_netw_r2}
\end{center}
\end{table}

\section{FIR-based analysis}
\label{sec:FIR_based}

To understand how the network structure changes at a smaller spatial scale, we further reconstructed the daily interaction networks corresponding to each FIR in the European airspace. To do so, each interaction is associated to the FIR in which it has taken place; the network is then formed by filtering only the relevant nodes. The same metrics of Tab. \ref{tab:metrics_def} are measured, and their predictability assessed against the total flown distance over each FIR.

\begin{figure}[tb]
\centering
\includegraphics[width=\linewidth]{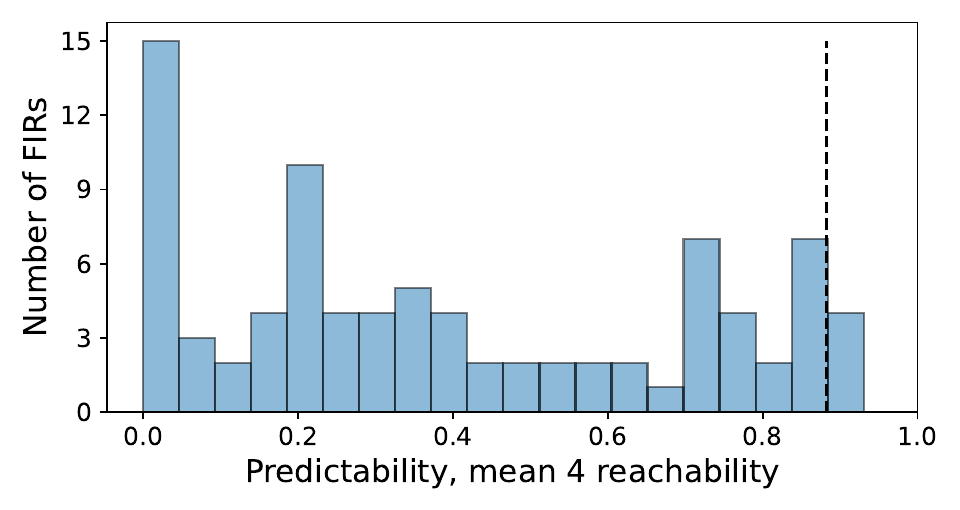}
\caption{Distribution of the predictability for the mean 4 reachability metric, for each of the analysed FIRs. The black dashed line represents the same metric for the complete European airspace - see Tab. \ref{tab:metrics_vs_distance_full_netw_r2}. }
\label{fig:r2_histogram}
\end{figure}

\begin{figure*}[tb]
\centering
\includegraphics[width=\textwidth]{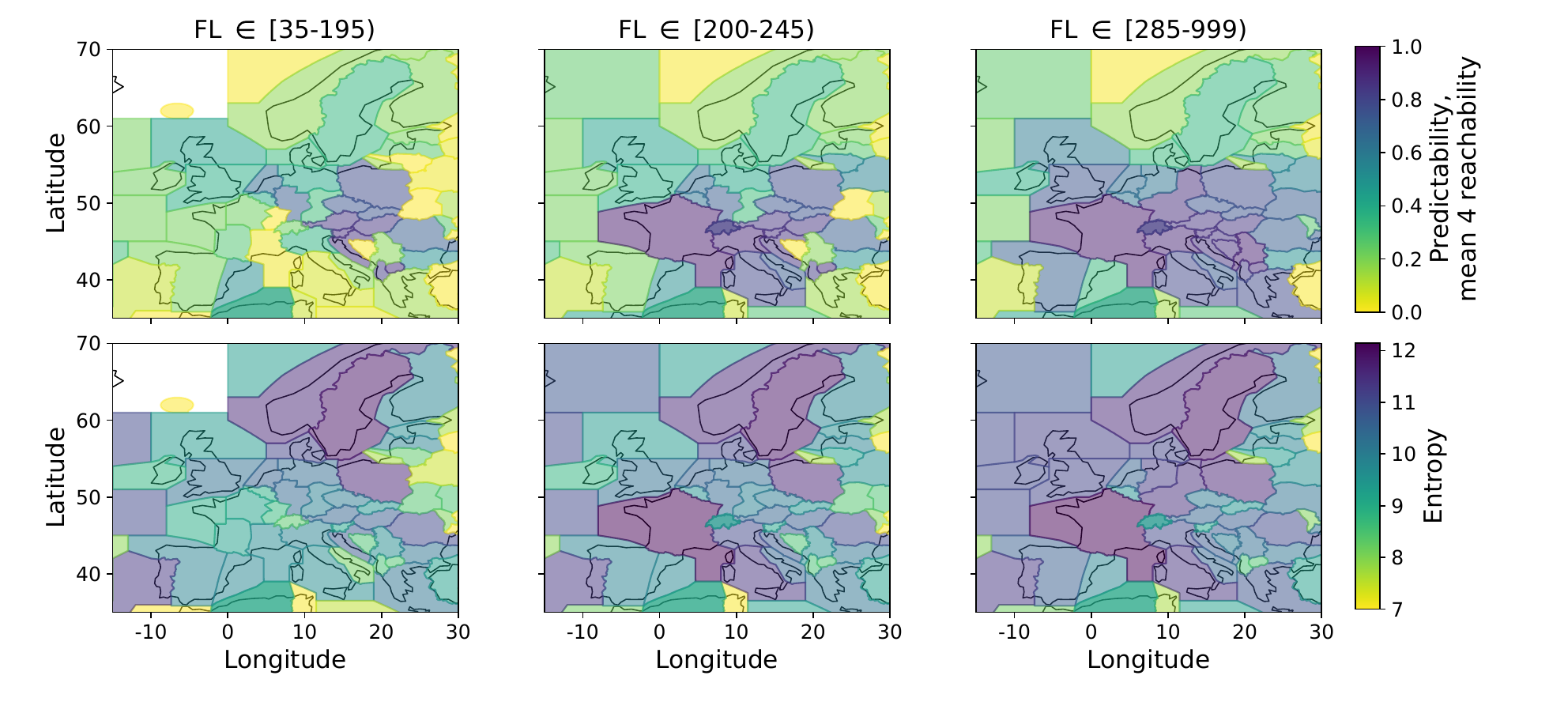}
\caption{Spatial representations. (Top) Predictability for each FIR, according to the mean 4 reachability. (Bottom) Entropy of the airspace usage in each FIR. Each panel represents the results for a range of FLs, see top labels. White areas correspond to FIRs with no significant data.
}
\label{fig:r2_and_entropy_map}
\end{figure*}

An example for the mean 4 reachability can be seen in Fig. \ref{fig:r2_histogram}, specifically depicting the probability distribution of the obtained predictability. It can be appreciated that most FIRs have a lower predictability, when compared to what obtained for the global airspace (see vertical dashed black line). A similar trend was found for the other topological metrics, not reported here due to space limitations. Two conclusions can here be drawn: (i) individual FIRs have a generally lower predictability than the global airspace, suggesting that this property is lost at a micro-scale; and (ii) there is a large variability in the predictability across FIRs.
When the predictability is represented in a spatial map, see top panels of Fig. \ref{fig:r2_and_entropy_map}, two patterns seem to emerge: larger values can be observed in the central and southern parts of Europe, and at higher altitudes. 

This suggests two explanations. On the one hand, the fact that central regions of Europe are associated with a higher $R^2$ may be due to the higher amount of traffic in them present. A strategy similar to the previous one can be used to test this, specifically performing a linear regression between the predictability of each region, and the corresponding total flown distance. Results are reported in Tab. \ref{tab:r2_metrics_vs_entropies} (second and third column). All fits are statistically significant, and the flown distance is able to explain between $20\%$ and $25\%$ of the variability - with the only exception been the modularity, see last row.

\begin{table}[tb]
\caption{$R^2$ and $p$-value for the best linear fit between the predictability of each topological metric, and the total flown distance and entropy of the FIRs. }
\begin{center}
\begin{tabular}{|l|c|c|c|c|}
\hline
 &\multicolumn{2}{|c|}{\textbf{Flown distance}} & \multicolumn{2}{|c|}{\textbf{Norm. entropy}} \\
\cline{2-5} 
\textbf{Metric} & $\mathbf{\mathit{R^2}}$& $\mathbf{\mathit{p}}$\textbf{\textit{-value}}  & $\mathbf{\mathit{R^2}}$& $\mathbf{\mathit{p}}$\textbf{\textit{-value}} \\\hline
Degree entropy & $0.224$ & $4\cdot 10^{-6}$& $0.609$ & $7\cdot 10^{-19}$ \\\hline
Isolated nodes&  $0.235$ & $2\cdot 10^{-6}$ & $0.614$& $4\cdot 10^{-19}$ \\\hline
Weak giant cluster size & $0.341$& $3\cdot 10^{-9}$ & $0.349$ & $2\cdot 10^{-9}$ \\\hline
Efficiency &  $0.334$& $5\cdot 10^{-9}$ & $0.303$ & $4\cdot10^{-8}$ \\\hline
Mean 4 reachability&  $0.225$& $3\cdot 10^{-6}$ & $0.557$ & $1\cdot 10^{-16}$ \\\hline
Reachability modularity & $0.047$& $4\cdot 10^{-2}$& $0.139$&  $4\cdot 10^{-4}$ \\\hline

\end{tabular}
\label{tab:r2_metrics_vs_entropies}
\end{center}
\end{table}

On the other hand, the differences between lower and higher airspaces may be due to the different route structures in them present. In order to test this, we have calculated an entropy of the airspace usage. Each FIR has been divided in square cells of size $0.01^{\circ} \times 0.01^{\circ}$ (approximately 0.5NM $\times$ 0.5NM); the total number of times a flight has flown within each cell has then be calculated. These values have been interpreted as a probability, by normalising their sum to one, with the FIR's entropy being the Shannon's entropy of the associated probability distribution \cite{aczel1974shannon}. In order to avoid biases due to different FIR sizes, this value is finally normalised by the maximum entropy $\ln\left(N\right)$, where $N$ is the number of cells in the corresponding FIR.

As in the previous case, Tab. \ref{tab:r2_metrics_vs_entropies} reports the relationships between this normalised entropy and the predictability of the topological metrics, calculated through a linear regression; see also bottom panels of Fig. \ref{fig:r2_and_entropy_map} for a spatial representation. The relationship is here even clearer, with the entropy being able to explain more than half of the variability for three metrics.

\section{Spatial multiscale structure}
\label{sec:multiscale}

\begin{figure}[tb]
\centering
\includegraphics[width=\linewidth]{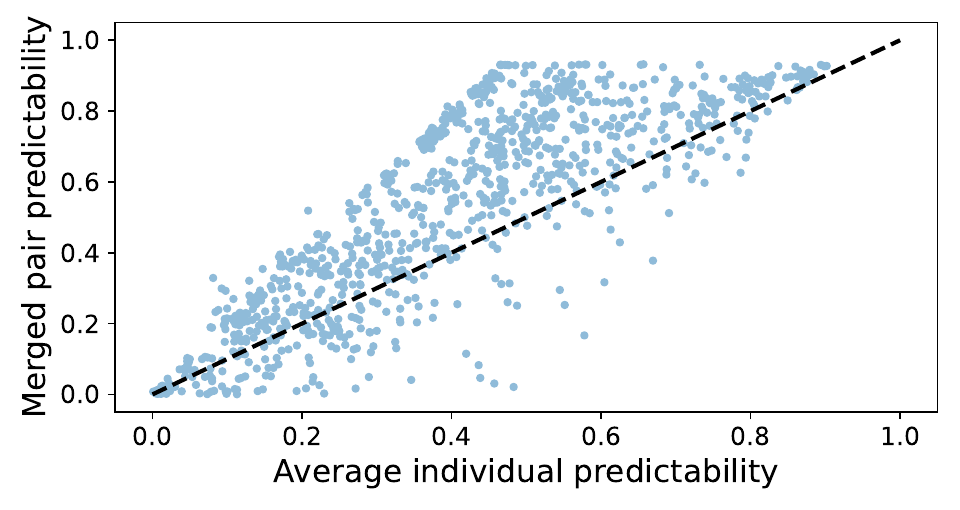}
\caption{Evolution of the predictability of merged pairs of FIRs, as a function of the average of the individual predictabilities; in both cases, for the mean 4 reachability. Each point represents a pair of FIRs, with $10^3$ random pairs being represented. For visual reference, the dashed black line indicates the main diagonal. }
\label{fig:multiscale}
\end{figure}

As a final topic, we briefly explore how the previously presented results scale spatially. We have seen how the predictability of topological metrics is related to the entropy of airspaces; yet, this alone cannot explain the high values observed for the complete airspace, as the entropy of the latter cannot be higher than that of its part - being entropy a subadditive property \cite{aczel1974shannon}. A multiscale analysis has then be designed, based on randomly joining pairs of FIRs, and calculating the resulting (joint) interaction network and the predictability of its metrics. We then plotted this merged predictability, as a function of the mean predictability of the two constituting FIRs - see Fig. \ref{fig:multiscale}. Notably, the large majority of points lay above the main diagonal, i.e. the former measure is larger than the latter; similar results, not reported here for space restrictions, were found for the other topological metrics. Some initial conclusions about this relationship will be drawn below.

\section{Discussion and future steps}
\label{sec:conclusions}

Leveraging previous results on the creation and analysis of complex networks representing interactions between aircraft \cite{lopez2023, lopez2024}, we here reported an analysis of their topological predictability and of the factors affecting it.
While only representing an initial investigation on the topic, the results here obtained allow drawing some interesting conclusions. The topological predictability seems to be driven by two main factors: the amount of traffic and its spatial organisation. As seen in Tab. \ref{tab:r2_metrics_vs_entropies} and Fig. \ref{fig:r2_and_entropy_map}, such predictability is higher when flights are more uniformly distributed throughout the airspace; and this is especially evident at high altitudes and in the central part of Europe, where airways create a complex and dense mesh \cite{esteve2023structure}.
Note that advanced operational concepts, like the free routing in the upper airspace \cite{nava2018study}, may contribute to this result; but are not the cause of it, as we previously showed \cite{lopez2024}, as they minimally impact the structure of the network.
We hypothesise that the opposite is happening near major airports, where traffic is funneled into arrival (and departure) procedures that have a single final (starting) point, and hence where aircraft are closely packed together.

From an operational viewpoint, this could have some relevant implications. Insofar the objective of reconstructing these networks is the simplification of the macroscopic structures created by interacting aircraft, the entropy of the airspace utilisation represents an upper limit of what can be achieved. Topological metrics are more stable and predictable in highly transited and homogeneously used airspaces; any intervention therein, e.g. by rerouting or delaying specific flights, will thus have a minor impact. Notably, the entropy will be maximal in free-flight scenarios, which correspond to having a virtual airway for each flight; hence, the rigidity of the interaction structures will also be maximal. While we have previously shown that aircraft following geodesic routes do not create more complex structures \cite{lopez2024}, this may come at the cost of a smaller margin for improvement.

Results of Sec. \ref{sec:multiscale} also indicate that the topological predictability is not a subadditive property: the predictability of an airspace is higher than the sum of the predictability of its composing parts. On the one hand, this may be the result of the resolution limits  of the topological metrics here considered, a problem well known in the case of the modularity \cite{arenas2008analysis}. 
On the other hand, we speculate that this is also reflecting the way the airspace is designed, with FIRs being a natural spatial scale at which flights are organised. This could be tested by analysing the structures emerging within individual sectors, i.e. the smaller organisational scale; and further opens door to evaluate whether intermediate spatial scales are relevant to understand interaction propagations. 

Finally, while previous analyses indicated that the structure of the networks is fairly stable against changes in the parameters of the reconstruction process \cite{lopez2023}, the latter ones may have an impact in the predictability. Results here presented could therefore be analysed under variations of e.g. the separation thresholds, or when interactions are filtered according to the approach angle or the relative speed of the aircraft.



\bibliographystyle{IEEEtran}
\bibliography{Bibliography.bib}

\end{document}